\documentclass[usenatbib]{mn2e}
\usepackage{amsmath, graphicx, times, gensymb}
\def\apj{{ApJ}}                 
\def\apjl{{ApJ}}                
\def\apjs{{ApJS}}               
\def\aap{{A\&A}}                

\def\mnras{{MNRAS}}             
\def\prd{{Phys.~Rev.~D}}        
\def\pasj{{PASJ}}               

\def\physrep{{Phys.~Rep.}}   

\title[Redshift Distortion Free 3PCF in the nonlinear regime]
{A redshift distortion free correlation function at third order in the nonlinear regime}

\begin{document}

\author[Meng, Pan, Szapudi \& Feng]{Kelai Meng$^1$, Jun Pan$^{1,2}$\thanks{jpan@pmo.ac.cn},  
Istv\'{a}n Szapudi$^2$ and Longlong Feng$^1$\\
$^1$Purple Mountain Observatory, 2 West Beijing Rd., Nanjing 210008, P. R. China\\
$^2$Institute for Astronomy, University of Hawaii, 2680 Woodlawn Dr., HI 96822, USA}

\maketitle

\begin{abstract}

The zeroth-order component of the cosine expansion of the
projected three-point correlation function is proposed for clustering 
analysis of cosmic large scale structure. These functions are third order
statistics but can be measured similarly to the projected 
two-point correlations. Numerical experiments 
with N-body simulations
indicate that the advocated statistics are redshift distortion free
within $10\%$ in the non-linear regime on scales 
$\sim 0.2-10 h^{-1}$Mpc. 
Halo model prediction of the zeroth-order component of the projected three-point 
correlation function agrees with simulations within $\sim 10\%$.
This lays the ground work for using these functions to perform joint
analyses with the projected two-point correlation functions,
exploring galaxy clustering properties in the framework of the halo model and 
relevant extensions.
\end{abstract}


\section{Introduction}
Observed large scale structure in the Universe is generally 
conjectured to arise from Gaussian initial condition or nearly so;
the rather high level non-Gaussianity at present is due 
to the action of gravitational force and gas
physics. The three-point correlation 
function (3PCF) is of the lowest order among correlation functions 
capable of probing such non-Gaussianity. With the
recent increase of interest and the corresponding attempts
to extract more information about structure formation processes and
primordial non-Gaussianity from fine clustering patterns of galaxies,
the 3PCF (or its counterpart in Fourier space, bispectrum) has 
attracted much attention in recent years 
\citep[e.g][]{KayoEtal2004, NicholEtal2006, 
SmithEtal2008, JeongKomatsu2009, Sefusatti2009}. 

However, 3PCF is well known for its low return of investment compared with 
the two-point correlation function (2PCF). 
One major obstacle hindering the interpretation and consequently
the application of 3PCF is the redshift distortion induced by the peculiar 
velocities of galaxies. Although effects of redshift distortion on 2PCF 
(or power spectrum)
are not yet well understood analytically \citep[e.g.][]{Scoccimarro2004},
approximations by incorporating pairwise velocity distribution have been 
proposed, validated and applied successfully to statistical analyses
\citep{Peebles1980, DavisPeebles1983, White2001, Seljak2001, KangEtal2002, 
Tinker2007, SmithEtal2008}. In
the case of 3PCF (or bispectrum) analogous approach
would involve higher order statistics of peculiar velocities. 
The complicated entanglement
of redshift distortions with nonlinear gravitational dynamics and
nonlinear biasing renders theoretical prediction extremely difficult 
in configuration space. In Fourier space and with the distant observer
approximation, prediction of the
bispectrum in redshift space in various perturbative and empirical schemes
has been moderately successful, although none have been able to show
satisfactory agreement with simulations 
\citep{MatsubaraSuto1994, HivonEtal1995, VerdeEtal1998, ScoccimarroEtal1999}.
The mostly accurate model to date appears to be the work of 
\citet{SmithEtal2008}, a halo model extension implemented with 
higher order perturbation theory.

One can eliminate the complexity of redshift distortion 
with projection of the correlation functions upon the plane perpendicular
to the line-of-sight (LOS). Projected correlation functions are obtained by 
integrating over the anisotropic correlation functions along LOS, 
which effectively removes redshift distortions if the
conservation of total number of galaxy pairs and triplets along LOS can be 
satisfied. Since thickness of a realistic sample is finite, galaxies 
near radial edges could enter or leave the sample space by their apparent
movement due to peculiar velocity, such conservation is only approximately 
achieved if the sample is shallow, or redshifts are photometric. Violation of
the conservation condition may bring non-negligible systematical bias on large scales
\citep{NockEtal2010}. Nevertheless, this is not a problem for most modern
spectroscopic galaxy samples, and the bias actually can be minimized by 
careful design of estimation methodology.

In comparison with the projected 2PCF that has been widely used to investigate 
clustering dependence on galaxy intrinsic properties, evolution history and
environment and to distinguish cosmological models
\citep[e.g.][]{HawkinsEtal2003, ZhengEtal2007, BaldaufEtal2010, ZehaviEtal2010},
exploration and application of the projected 3PCF has been limited
in the literature 
\citep{JingBoerner1998, JingBorner2004a, Zheng2004, McBrideEtal2010}. 
Lack of accurate theoretical models of 3PCF
prevents proper interpretation of measurements. 
\citet{ScoccimarroCouchman2001} offered a phenomenological model based on
hyper-extended perturbation theory for the bispectrum in the nonlinear regime.
Their fitting formula is accurate on smaller scales
but in the weakly and mildly nonlinear regimes it is improved upon by the 
empirical model of 
\citet{PanColesSzapudi2007}. Both fail on very small scales, and neither 
can capture the signal of baryonic oscillation in bispectrum 
appropriately \citep{SefusattiEtal2010}. 
The approach of halo model appears more promising, as it can reproduce most 
measurements in simulations for the bispectrum
\citep[e.g.][]{MaFry2000a, MaFry2000b, ScoccimarroEtal2001, 
SmithWattsSheth2006, SmithEtal2008} and the 3PCF in configuration space 
\citep[e.g.][]{TakadaJain2003, WangEtal2004, FosalbaPanSzapudi2005}. In spite
of disagreement with simulations for some configurations of 3PCF,
the halo model is still more attractive than the phenomenological
models for its clean and physically motivated parametrization to
galaxy biasing through e.g. the machinery of
the halo occupation distribution \citep[HOD,][]{BerlindWeinberg2002}.

Another reason for the scarce exploration of projected 3PCF is 
the complexity of estimation.
Computational requirement of 3PCF is demanding for currently available 
computers when millions of points are typical. The
additional task of decomposing the separations among three points for
projected 3PCF adds to the CPU load. Furthermore, the 3PCF is already
more prone to Poisson noise than the 2PCF, 
and typical bin width of scales for projected 
3PCF is even smaller than for the normal 3PCF. 
In order to suppress discreteness effects for a
reliable estimation, a high number density of points in the sample is crucial,
but often unrealistic for real surveys.

By analogy to the monople of 3PCF advocated by 
\citet{PanSzapudi2005a, PanSzapudi2005b}, 
we show that a third-order statistical function similar to the angular 
average of
the projected 3PCF is redshift distortion free and relatively easy to
estimate and model theoretically. In the next section, the definitions, 
and relation with
3PCF together with estimation algorithm is described. Section 3 presents 
numerical properties of the new statistical measure while in section 4 we demonstrate
the consistency of halo models to simulations of the new function. Summary and
discussion are in the last section.

\section{projected three-point correlation function and its zeroth-order component}
Let ${\bf r}={\bf x}_2-{\bf x}_1$ be the vector pointing to a point at position
${\bf x}_2$ from point at ${\bf x}_1$, the vector can be decomposed
to two components, separation along the line-of-sight (LOS) $\pi=r\mu$ with
$\mu$ being the cosine of the angle between the LOS and ${\bf r}$, 
and separation perpendicular to LOS $\sigma=(r^2-\pi^2)^{1/2}$, then 
we have the anisotropic 2PCF $\xi(\sigma, \pi)$ and so the 
projected 2PCF
\begin{equation}
\begin{aligned}
\Xi(\sigma)& \equiv \int_{-\infty}^{+\infty} \xi(\sigma, \pi) {\rm d}\pi\ \\
=& 2\int_{\sigma}^{+\infty}
\frac{r\xi(r){\rm d}r}{\sqrt{r^2-\sigma^2}}
= 2\int_{\sigma}^{+\infty}
\frac{s\xi(s){\rm d}s}{\sqrt{s^2-\sigma^2}}\ ,
\end{aligned}
\end{equation}
where ${\bf s}$ is the separation vector between two points 
measured in redshift space and the last step comes from conservation of 
total number of pairs along LOS. Inversion of $\Xi(\sigma)$ could directly
render 2PCF $\xi(r)$ although inversion of such Abel integration is unstable 
mathematically \citep{DavisPeebles1983}.

Similarly, giving three points at ${\bf x}_1$, ${\bf x}_2$ and ${\bf x}_3$, 3PCF
$\zeta(r_1,r_2, r_3)$ is of the the triangle configuration 
with three separations 
${\bf r}_1={\bf x}_2-{\bf x}_1=(\sigma_1, \pi_1)$,
${\bf r}_2={\bf x}_3-{\bf x}_2=(\sigma_2, \pi_2)$ and 
${\bf r}_3={\bf x}_1-{\bf x}_3=(\sigma_3, \pi_3)$, decomposition
of the three separations bring up anisotropic 3PCF 
$\zeta(\sigma_{1,2,3};\pi_{1,2,3})$ with $\sum\pi_{1,2,3}=0$, and
the projected 3PCF is just defined as \citep{JingBoerner1998, JingBorner2004a, Zheng2004}
\begin{equation}
\begin{aligned}
Z( & \sigma_1, \sigma_2, \sigma_3) \equiv 
\int_{-\infty}^{+\infty}\int_{-\infty}^{+\infty} 
\zeta \left(\sigma_{1, 2, 3}; \pi_{1, 2}\right) 
{\rm d} \pi_1 {\rm d}\pi_2\\
&=2\int_{\sigma_1}^{+\infty} \int_{\sigma_2}^{+\infty} 
\frac{r_1 r_2\left[\zeta(r_1, r_2, r_3^+)+
\zeta(r_1, r_2,r_3^-)\right] }{
\sqrt{(r_1^2-\sigma_1^2)(r_2^2-\sigma_2^2)}}
{\rm d}r_1 {\rm d}r_2 \ ,
\end{aligned}
\end{equation}
in which
\begin{equation}
\begin{aligned}
r_3^+ & = \sqrt{\sigma_3^2+\left( |\pi_1| + |\pi_2| \right)^2}\\
r_3^- & = \sqrt{\sigma_3^2+\left( |\pi_1| - |\pi_2|\right)^2}\ .
\end{aligned}
\end{equation}

\citet{Szapudi2004a} pointed out that 3PCF can be expanded with 
Legendre polynomials $P_\ell$ to isolate part of the configuration dependence,
\begin{equation}
\begin{aligned}
& \zeta(r_1, r_2, \theta) =\sum_{\ell=0}^{\infty} \frac{2\ell+1}{4\pi}
\zeta_\ell(r_1, r_2)P_\ell(\cos\theta) \\
& \zeta_\ell(r_1, r_2) =2\pi \int_{-1}^1 \zeta(r_1, r_2, \theta) 
P_\ell(\cos\theta) {\rm d}\cos\theta\ ,
\end{aligned}
\end{equation}
in which $\cos\theta=-{\bf r}_1 \cdot {\bf r}_2/(r_1 r_2)$. In the
expansion the monople $\zeta_0$ is of particular interests for its
relatively simplicity in measurement and 
interpretation \citep{PanSzapudi2005a, PanSzapudi2005b}. One can
easily found that $\zeta_0$ is actually 
the spherical average of $\zeta$ in three-dimensional space
\begin{equation}
\frac{\zeta_0(r_1, r_2)}{4\pi}=
\frac{\int \zeta(r_1, r_2, \theta) 2\pi \sin\theta {\rm d}\theta}{4\pi} =
\frac{\int \zeta{\rm d}\Omega}{\int {\rm d}\Omega}\ ,
\end{equation}
which effectively becomes theoretical support to the estimator in
\citet{PanSzapudi2005b}.

In the same spirit, the projected 3PCF $Z$ also can be expanded but 
in a different treatment, the cosine Fourier transformation proposed 
by \citet{Zheng2004} and \citet{Szapudi2009} is the appropriate one since $Z$ is defined on a
two-dimensional plane which is perpendicular to LOS.
Angular averaging of $Z$ thus produces the zeroth-order component of the 
cosine expansion to $Z$ \citep{Zheng2004,Szapudi2009},
\begin{equation}
Z_0(\sigma_1, \sigma_2)=\frac{1}{2\pi}
\int_0^{2\pi} Z(\sigma_1, \sigma_2, \theta_p) {\rm d}\theta_p
\end{equation}
with $\theta_p=\cos^{-1} [ (\sigma_1^2+\sigma_2^2-
\sigma_3^2)/(2\sigma_1 \sigma_2 ) ]$, 
which is the object function that we focus on and actually is related 
to $\zeta$ by
\begin{equation}
\begin{aligned}
Z_0 & =  \frac{1}{2\pi}\int_0^{2\pi} {\rm d}\theta_p 
\int_{-\infty}^{+\infty}\int_{-\infty}^{+\infty}
\zeta \left(\sigma_1,\sigma_2,\theta_p; \pi_1, \pi_2 \right) 
{\rm d} \pi_1 {\rm d}\pi_2  \\
& = \int_{-\infty}^{+\infty} \int_{-\infty}^{+\infty} 
\widetilde{\zeta_0}(\sigma_1, \sigma_2, \pi_1, \pi_2)
{\rm d}\pi_1 {\rm d}\pi_2\ 
\end{aligned}
\label{eq:z0}
\end{equation}
where
\begin{equation}
\widetilde{\zeta_0}(\sigma_1, \sigma_2, \pi_1, \pi_2)
= \frac{1}{2\pi}\int_0^{2\pi}
\zeta \left(\sigma_1,\sigma_2,\theta_p; \pi_1, \pi_2\right) {\rm d}\theta_p\ .
\label{eq:az0}
\end{equation}
Note that $\widetilde{\zeta_0}$ and the monople of 3PCF $\zeta_0$ are not equal at all.

Theoretically if the nonlinear bispectrum is known, 
by the cosine transformation
\begin{equation}
\begin{aligned}
& B(k_1, k_2, \phi)=
\sum_{n=-\infty}^{+\infty} B_n(k_1, k_2) \cos(n\phi) \\
& B_n(k_1, k_2)=\frac{1}{2\pi}\int_0^{2\pi} B(k_1, k_2, \phi) \cos(n \phi) {\rm d}\phi\ ,
\end{aligned}
\end{equation}
it is fairly straightforward to compute $Z_0$ \citep{Zheng2004}
\begin{equation}
\begin{aligned}
Z_0(\sigma_1,\sigma_2)= \frac{1}{(2\pi)^2} & 
\int_0^{+\infty} \int_0^{+\infty} B_0(k_1, k_2) \\
\times &J_0(k_1\sigma_1) J_0(k_2\sigma_2) k_1 k_2
{\rm d} k_1 {\rm d} k_2 \ ,
\end{aligned}
\label{eq:b0z0}
\end{equation}
where $J_0$ is the zero-order Bessel function of the first kind.
An immediate fact is that $Z_0$ only requires good approximation to the 
zeroth-order component of the nonlinear bispectrum, which simplifies 
theoretical development.

At a first glance it seems that it is not useful to 
invoke $Z_0$, since $Z$ contains more information,
the former erases the angular dependence completely through 
averaging. However, by smoothing $Z_0$ suffers much less from
shot-noise than $Z$, i.e. has smaller variance, which is a celebrated
property particularly when sample is not of high number density. 
More important, as we will see in next section, $Z_0$ can be easily estimated
with the common procedure for anisotropic 2PCF after some minor
modification. The savings in computing time, proportional
to the number of galaxies, is tremendous compared with
calculating the projected 3PCF of the full configuration.

\section{estimation and numerical test}
\subsection{Estimator}
Estimation of the zeroth-order component of the projected 3PCF
is based on Eqs.~\ref{eq:z0} and \ref{eq:az0}. Eq.~\ref{eq:az0}
indicates that $\widetilde{\zeta_0}$ can be measured with the same estimator
of $\zeta_0$ as in \citet{PanSzapudi2005b}, taking the same form of
the one in \citet{SzapudiSzalay1998}, 
\begin{equation}
\widetilde{\zeta_0} =\frac{DDD-3DDR+3DRR-RRR}{RRR}\ ,
\end{equation}
grouped symbols of $D$ and $R$ refer to various normalized number counts 
of triplets similar to what is in \citet{PanSzapudi2005b}, difference 
is that $\widetilde{\zeta_0}$ is estimated in bins of both $\sigma$ and $\pi$. 
Explicitly, if scale bins are linear, given two vector 
bins ${\bf r}_{jk}=(\sigma_j, \pi_k)$ and
${\bf r}_{j'k'}=(\sigma_{j'}, \pi_{k'})$ , with
$\sigma_{jk}$ in $(\sigma_{jk}-\Delta \sigma/2, \sigma_{jk} +\Delta\sigma/2)$, 
and $\pi_{jk}$ in $(\pi_{jk}-\Delta\pi/2, \pi_{jk}+\Delta\pi/2)$, 
as an example, the $DDD$ is obtained through
\begin{equation}
DDD=\left\{ \begin{array}{cc} \frac{\sum_{i=1}^{N_g} 
n_i({\bf r}_{jk}) \left[n_i({\bf r}_{j'k'})-1 \right]}{N_g (N_g-1) (N_g-2)}\ , 
& \textrm{if}\ {\bf r}_{jk}={\bf r}_{j'k'} \\
\frac{\sum_{i=1}^{N_g} n_i({\bf r}_{jk}) n_i({\bf r}_{j'k'})}
{N_g (N_g-1) (N_g-2)}\ , & \textrm{if}\ {\bf r}_{jk}\ne {\bf r}_{j'k'}
\end{array} \right. \ ,
\end{equation}
where $n_i$ is the number of neighbours to the center point counted
in the vector bin ${\bf r}_{jk}$.
Then by Eq.~\ref{eq:z0} integrating $\widetilde{\zeta_0}$ over $\pi_{k}$
and $\pi_{k'}$ yields estimation of $Z_0$. We have to address here
that unlike 3PCF, the estimator can not completely eliminate edge effects
for $\zeta_0$, $\widetilde{\zeta_0}$ and so $Z_0$, one needs to be cautious
when scales at probe is comparable to sample's characteristic size.

\subsection{Data preparation and estimation setup}
Since our goal is to provide a redshift distortion free third-order statistics, 
a key question is whether $Z_0$ measured in redshift space agrees with what we get 
in real space. In absence of accurate models about redshift distorted 3PCF, 
particularly in nonlinear regime, the best approach is to work with N-body simulation 
data directly. Two realizations of LCDM simulations run with Gadget2
\citep{Springel2005} were  analysed. Their cosmological parameters are 
taken from WMAP3 
fits \citep{SpergelEtal2007}, $\Omega_m=0.236$, $\Omega_\Lambda=0.764$, $h=0.73$ 
and $\sigma_8=0.74$. $512^3$ particles were evolved in both simulations, but one box 
size is $L=300h^{-1}$Mpc (box300) and the other is $L=600h^{-1}$Mpc (box600), the
force softening lengths are $12h^{-1}$kpc  and $24h^{-1}$kpc respectively. 
The $z=0$ output of box300 simulation and $z=0.09855$ output of box600 simulation 
were selected for our numerical experiment.

It is unpractical to use all particles in the simulations therefore
for each set of data we generate nine diluted samples for analysis to
control the amount of computation at a reasonable level; all results we present here 
are mean values of nine runs, and the actual scatter of different realizations
is very small. For box300 the number of randomly picked points 
is about $0.2\%$ of the total, while for box600 more than $\sim 600,000$ points are used. 
Several other samples diluted at different levels were also generated for 
consistency check. We find that sample dilution does affect our
estimation of $Z_0$ but mainly on very small scales, and that variance due to discreteness 
becomes larger with fewer points, as expected.

A common assumption about redshift distortions is the plane parallel
approximation (distant observer assumption), which assumes that the observer is 
very far away from the sample so that all lines-of-sight
from the observer to galaxies are parallel to each other. 
It simplifies calculation by reducing a 3D problem into 1D and indeed 
works well when the interested scale opens only a narrow angle to the observer. But 
the systematic bias introduced by the plane parallel
approximation turns out to be significant if the angle becomes wide. Theoretical
calculations and numerical measurements have shown that the deviation mainly
occurs at relatively large scales and could be more than $10\%$
\citep[e.g.][]{SzalayEtal1998, Scoccimarro2000, Szapudi2004b, CaiPan2007, PapaiSzapudi2008}.
To test the accuracy of the plane parallel approximation, two sets
of samples in redshift space are generated for the box300 data, one set takes
the distant observer assumption while the other mimics realistic samples by placing
an observer at distance of $100h^{-1}$Mpc to the nearest surface of the sample.
One has to bear in mind that the two redshift distortion 
scenarios differ not only in sample construction but also the way of decomposing
the separation ${\bf r}$ into $(\sigma, \pi)$ amid measurement. The output of box600
simulation we used is of $z\approx 0.1$, the plane parallel approximation
is sufficient for most analysis if interested scales are less than $\sim 50h^{-1}$Mpc.
We also noticed that applying periodic boundary condition or simply throwing away
those points shifted out of box by peculiar velocity makes
little difference for the final $Z_0$, which effectively eliminates the concern
of \citet{NockEtal2010}. 

During our estimation the $\sigma$ bins are set 
logarithmic with $\Delta\log \sigma=\log 1.4$,  $\pi$ bins are linear with 
$\Delta\pi=3, 5h^{-1}$Mpc for the box300 and the box600 respectively. Caution
must be taken about the bin width which shall not be too wide to
degrade the accuracy too much, while it shall be large enough to 
achieve $DDD>0$ for even the narrowest bin. Experience shows that
normally $DDD > 100$ is good enough to give reliable estimation
at our accuracy goal of 10 \%.

As we do not have multiple realizations to produce error bars, for each simulation
data set we split the sample into eight sub-volume boxes in half size, then the
scatter of measurements in these eight sub-volume boxes are taken as an
estimate of the  variance. 

\subsection{Finite integration range along LOS}

\begin{figure*}
\resizebox{\hsize}{!}{
\includegraphics{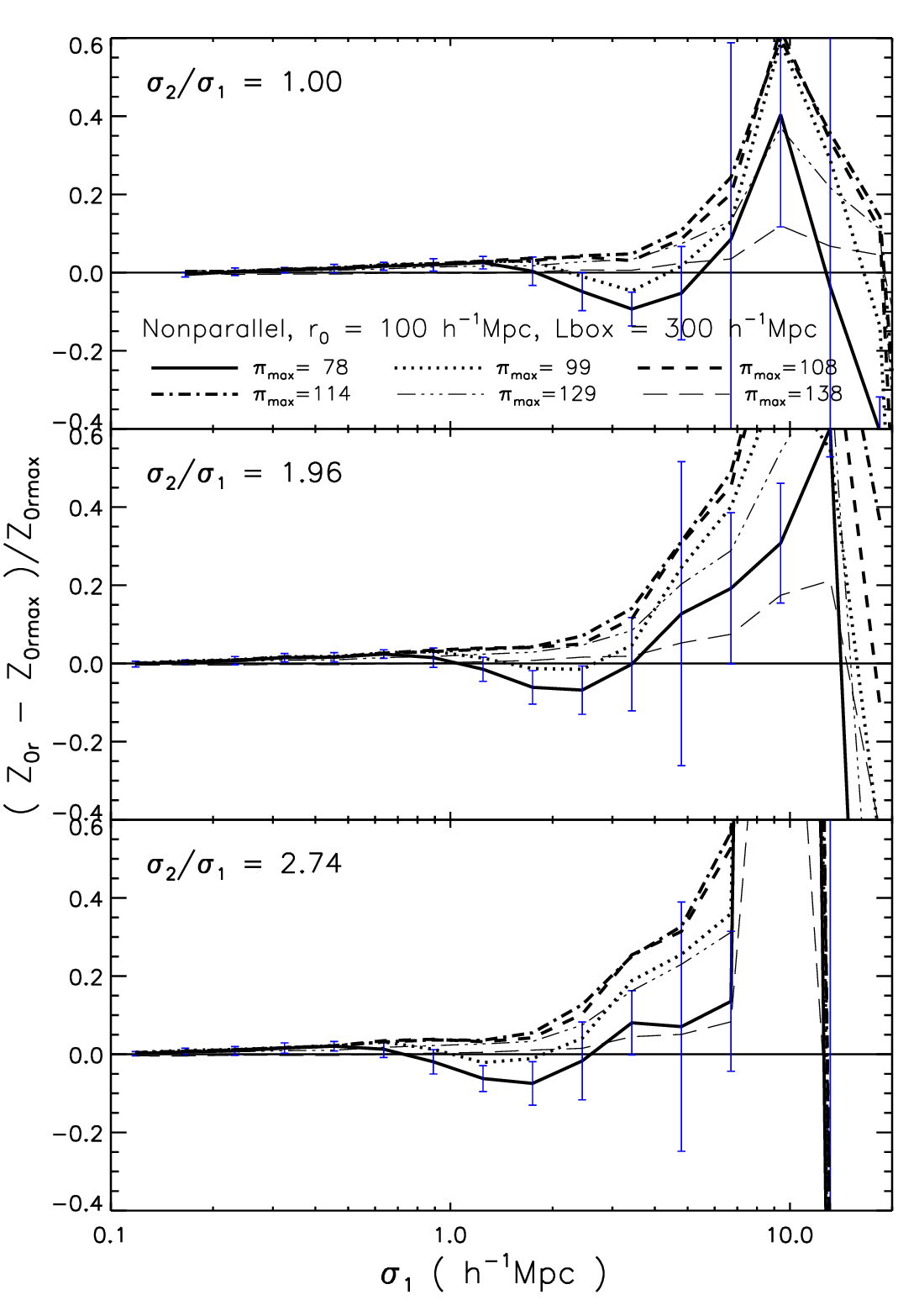}
\includegraphics{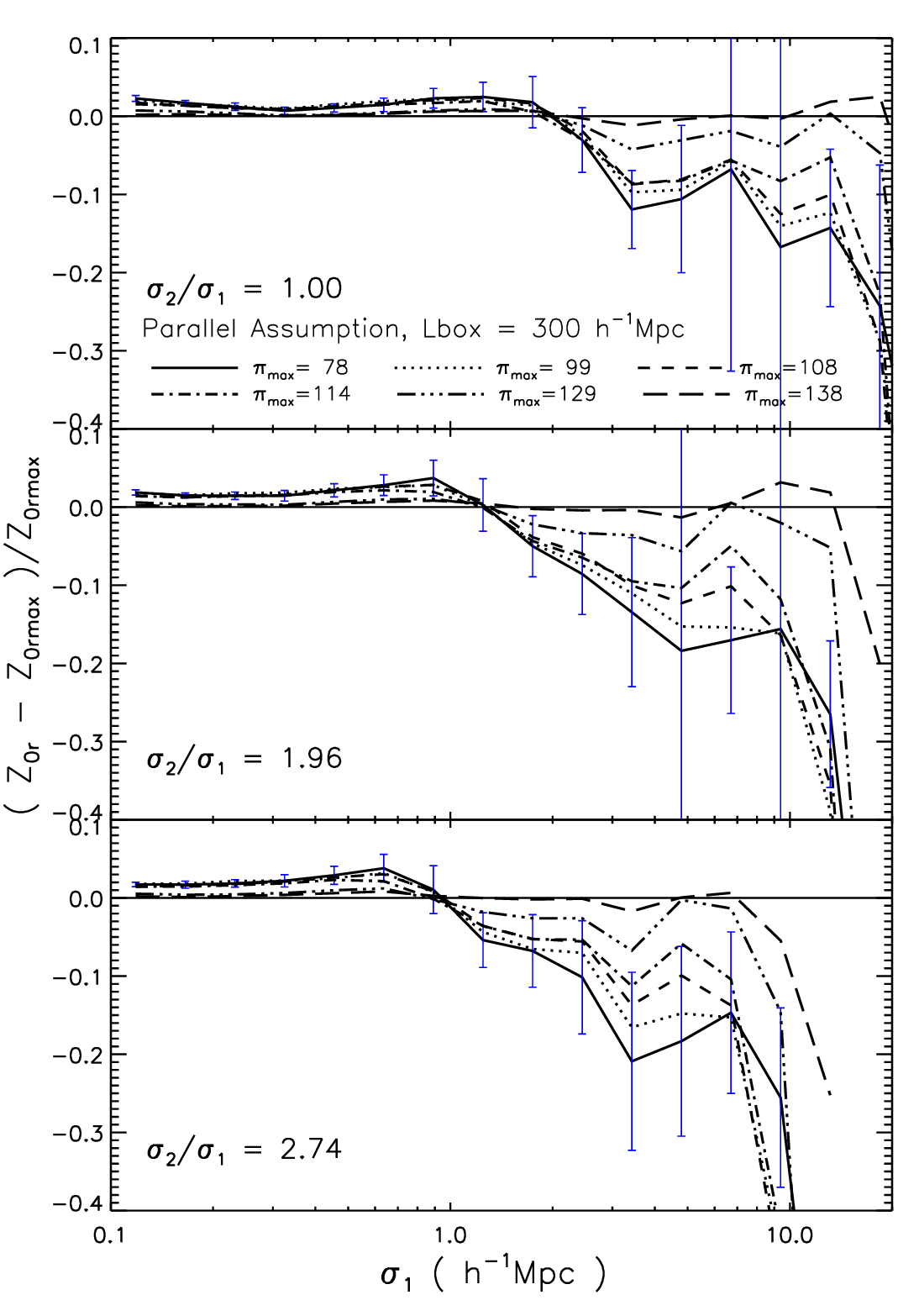}
\includegraphics{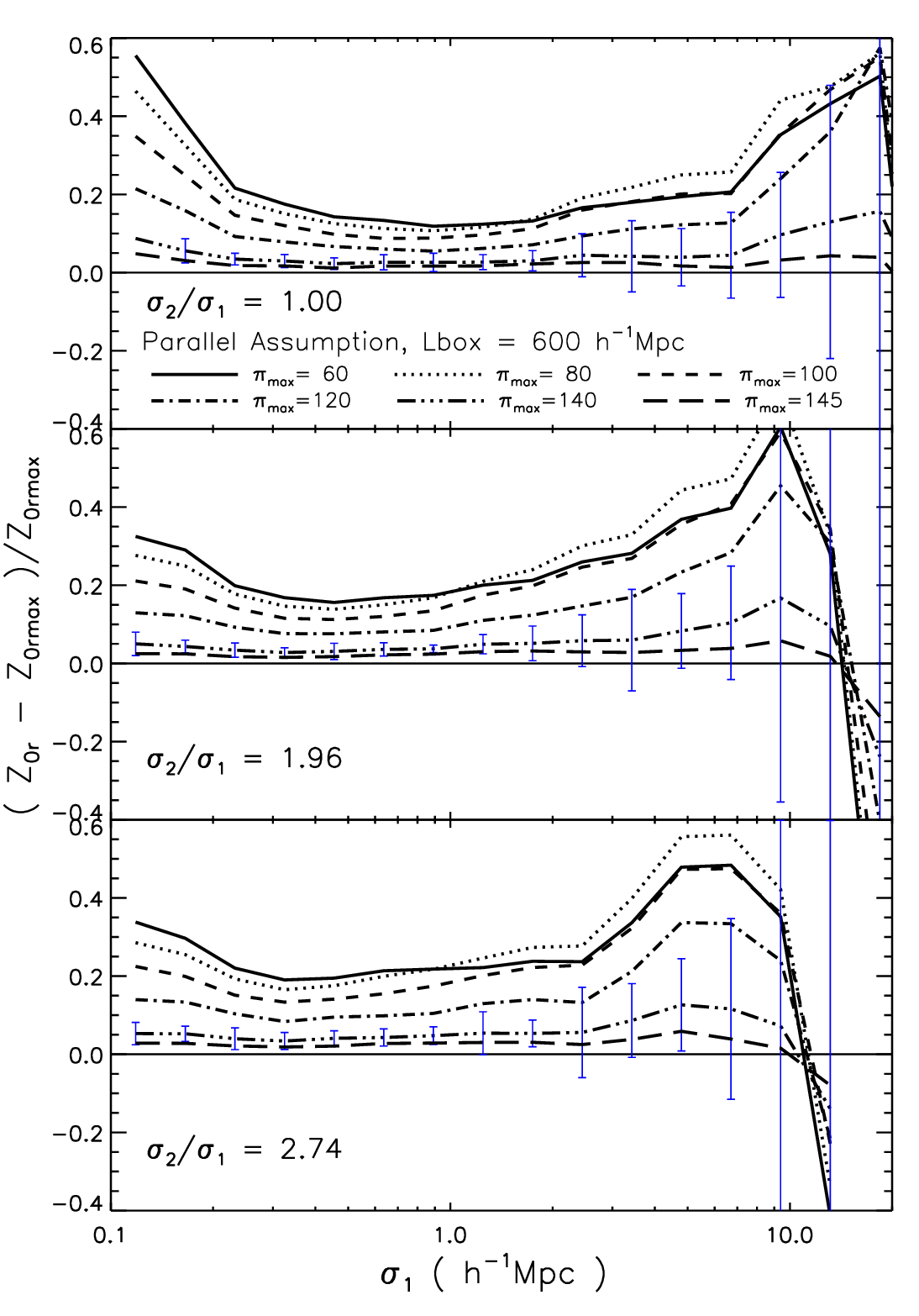}}
\caption{Fractional change of $Z_0(\sigma_1, \sigma_2)$ 
with $\pi_{max}$ compared to the one measured with
largest $\pi_{max}$. Three classes of configuration of 
$\sigma_2/\sigma_1$ of $Z_0$ are presented.
$Z_{0r}$ measured in real space with 
$\pi_{max}$ as specified in the legend, $Z_{0rmax}$ is estimated with 
the largest $\pi_{max}=150h^{-1}$Mpc. 
Plane parallel approximation in real space 
means merely the decomposition of ${\bf r}$ into $(\sigma, \pi)$ 
using parallel LOS, while nonparallel corresponds to an
external observer at the distance of $100h^{-1}$Mpc to the bottom of 
the simulation box. Errorbars of $\pi_{max}=78h^{-1}$Mpc for box300, 
$\pi_{max}=120h^{-1}$Mpc for box600 are plotted to show the uncertainty.}
\label{fig:pimax}
\end{figure*}

The integration range along LOS to 
have $Z_0$ from $\widetilde{\zeta_0}$ 
in Eq.~(\ref{eq:z0}) should be $(-\infty, +\infty)$ to guarantee
conservation of triplets along LOS. 
However, one can not integrate infinite scales due to finite
radial thickness of realistic samples, so there is always a finite upper 
limit of $\pi$ to the integration. 
Our measurements thus correspond to
\begin{equation}
\hat{Z}_0=\int_{-\pi_{max}}^{+\pi_{max}}\int_{-\pi_{max}}^{+\pi_{max}}
\widetilde{\zeta_0}{\rm d}\pi_1{\rm d}\pi_2 
=\sum_{i, j}\widetilde{\zeta_0}\Delta\pi_i\Delta\pi_j \ .
\label{eq:sum}
\end{equation}
Let subscript $r$ denote quantities in real space and $s$ for those 
in redshift space. The practical limitation 
certainly introduces systematic bias, henceforth mathematically
$\hat{Z}_{0s} \ne \hat{Z}_{0r}\ne Z_0$. What we hope is that we can
find a $\pi_{max}$ so that the contribution from $\pi$ larger than that
is negligible at a given tolerance. In our test runs 
we found that the largest $\pi_{max}$ permitted is around 
$1/4-1/2$ of the box size. If larger scale is used, the estimator of 
$Z_0$ suffers greatly of finite-volume effects. 
The same problem is present when estimating projected 2PCF, and 
normally it is agreed that $\pi_{max}\sim 40-70h^{-1}$Mpc is sufficient
to give stable results at small $\sigma$ of less than $\sim 20-30h^{-1}$Mpc, 
but may not be enough for measurement on larger scales 
\citep[see][and references there in]{BaldaufEtal2010}.

Figure~\ref{fig:pimax} presents the convergence of measurements
with changing $\pi_{max}$. It displays the fractional differences of 
$Z_0$ compared to that calculated from the largest $\pi_{max}$ 
allowed by the geometry of sample. Samples
used in this test are all in real space, but for box300 data we decompose
scales by LOS in two ways: plane parallel approximation and 
wide angle treatment.

For box300 at scales $\sigma< 1h^{-1}$Mpc $Z_0$ is extremely stable 
against different choices of $\pi_{max}$, independent of the scheme of 
scale decomposition, but at larger $\sigma$ scales the influence 
of those large $\pi$ becomes more and more evident. It appears that 
in the wide angle treatment $Z_0$ actually increases with $\pi_{max}$ 
up to $\sim 110h^{-1}$Mpc and then falls down when
further enlarging $\pi_{max}$, while in the plane parallel assumption 
$Z_0$ monotonically rises with larger $\pi_{max}$. The results from
box600 are somewhat different: $Z_0$ decreases with increasing 
$\pi_{max}$ on all $\sigma$ scales. Additional numerical experiments with 
the box600 data revealed that this behaviour is largely caused by the
dilution of the original data: a denser sample has
less variation against the choice of $\pi_{max}$ when $\sigma < 1h^{-1}$Mpc.

We conclude that for an overall precision target
of $\sim 10\%$ for $\sigma$ scales below $10h^{-1}$Mpc, 
$\pi_{max}\approx 120h^{-1}$Mpc suffices. This is
much larger than customary for the projected 2PCF. Note that the sharp 
break down of convergence at scales $\sigma\sim 10-20h^{-1}$Mpc
appears to be a numerical artifacts where $Z_0$ quickly approaches zero.

\subsection{Redshift space {\it versus} real space}

\begin{figure*}
\resizebox{\hsize}{!}{
\includegraphics{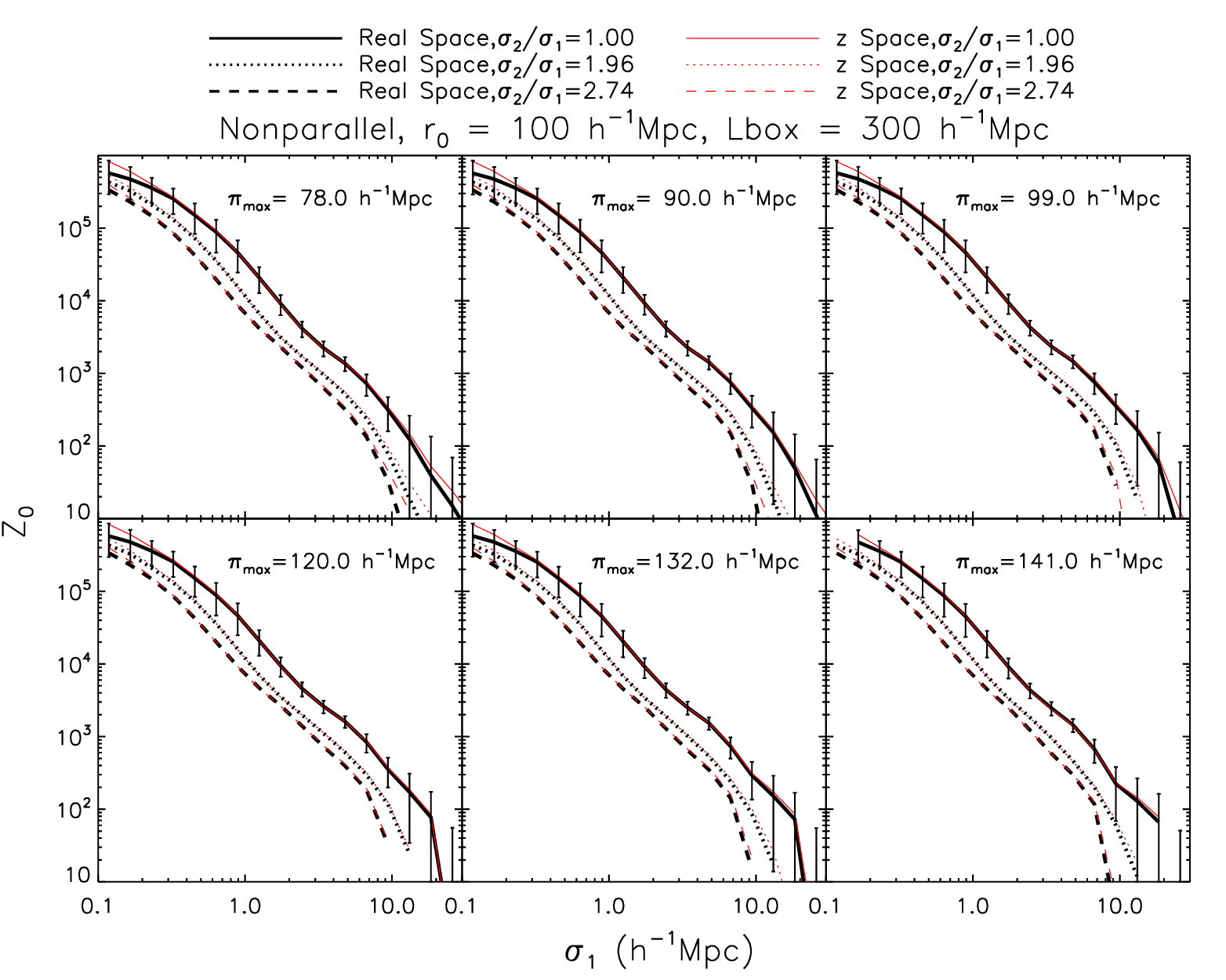}
\includegraphics{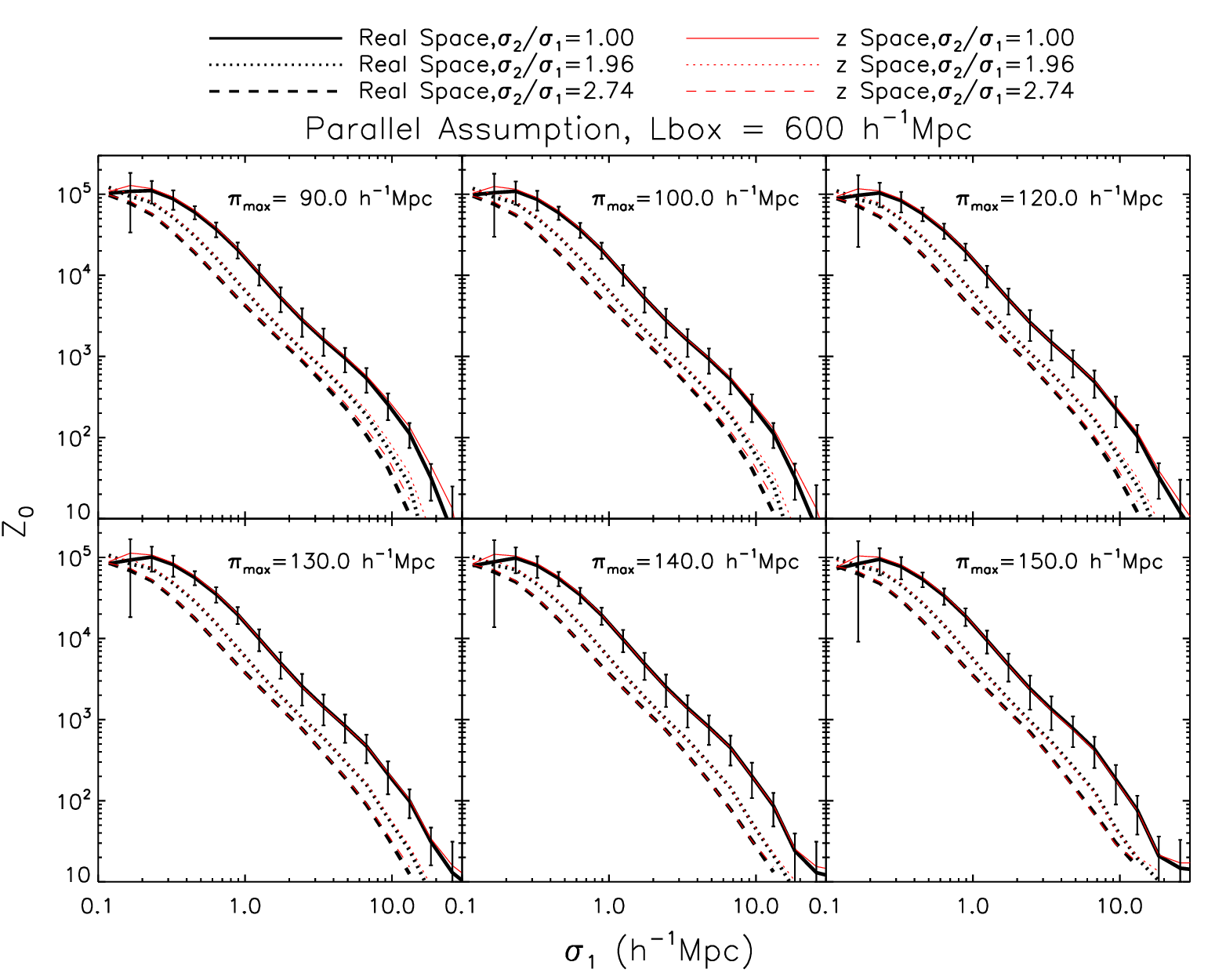}}
\caption{$Z_0$ measured in real space (thick lines) and that in redshift space (thin lines)
for different $\sigma_2/\sigma_1$ and $\pi_{max}$. Errorbars are of measurements of
configuration $\sigma_2/\sigma_1=1$ in real space. The results 
not shown for box300 under
plane parallel approximation are similar to the non-parallel case.}
\label{fig:z0rs}
\end{figure*}

\begin{figure*}
\resizebox{\hsize}{!}{
\includegraphics{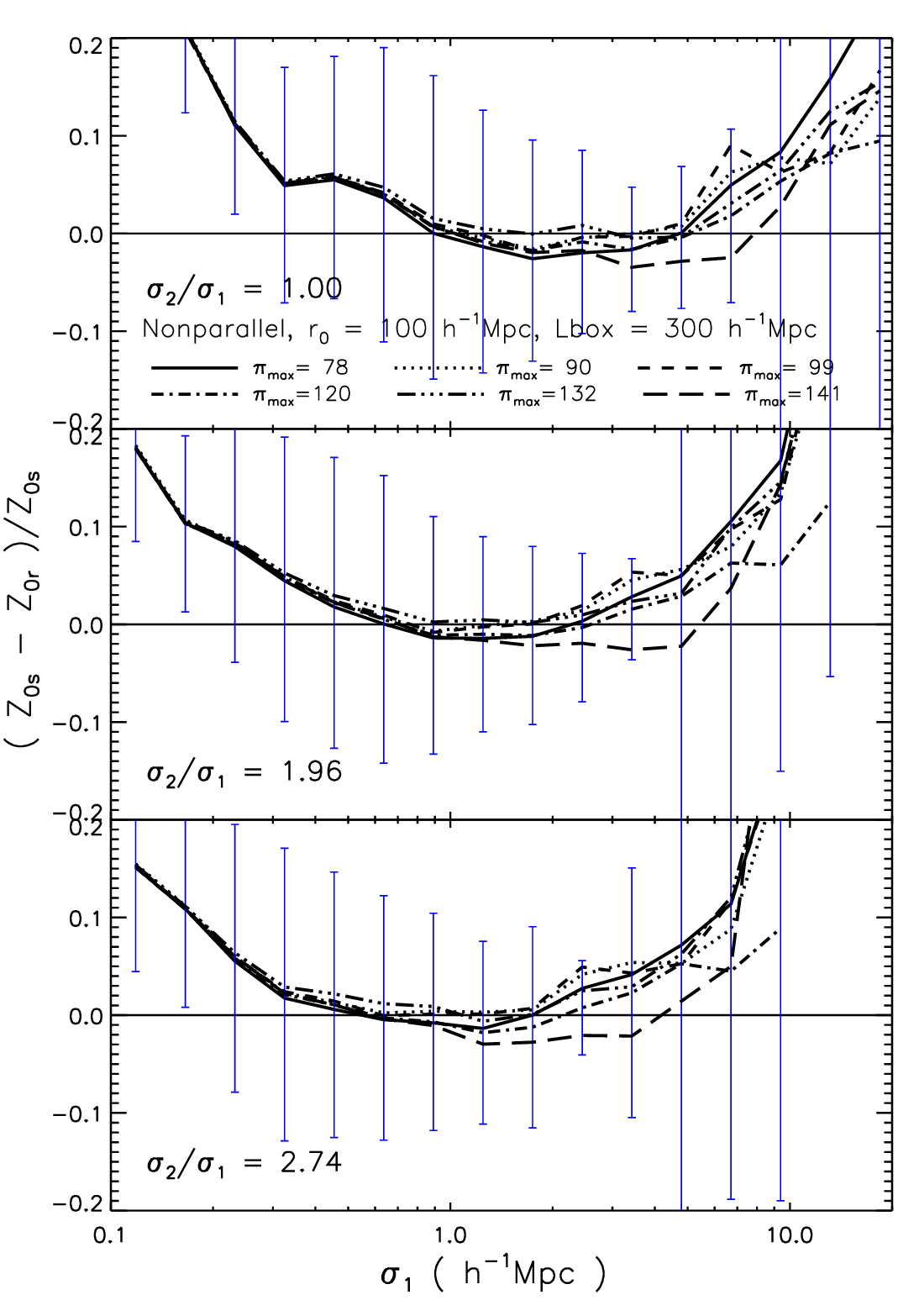}
\includegraphics{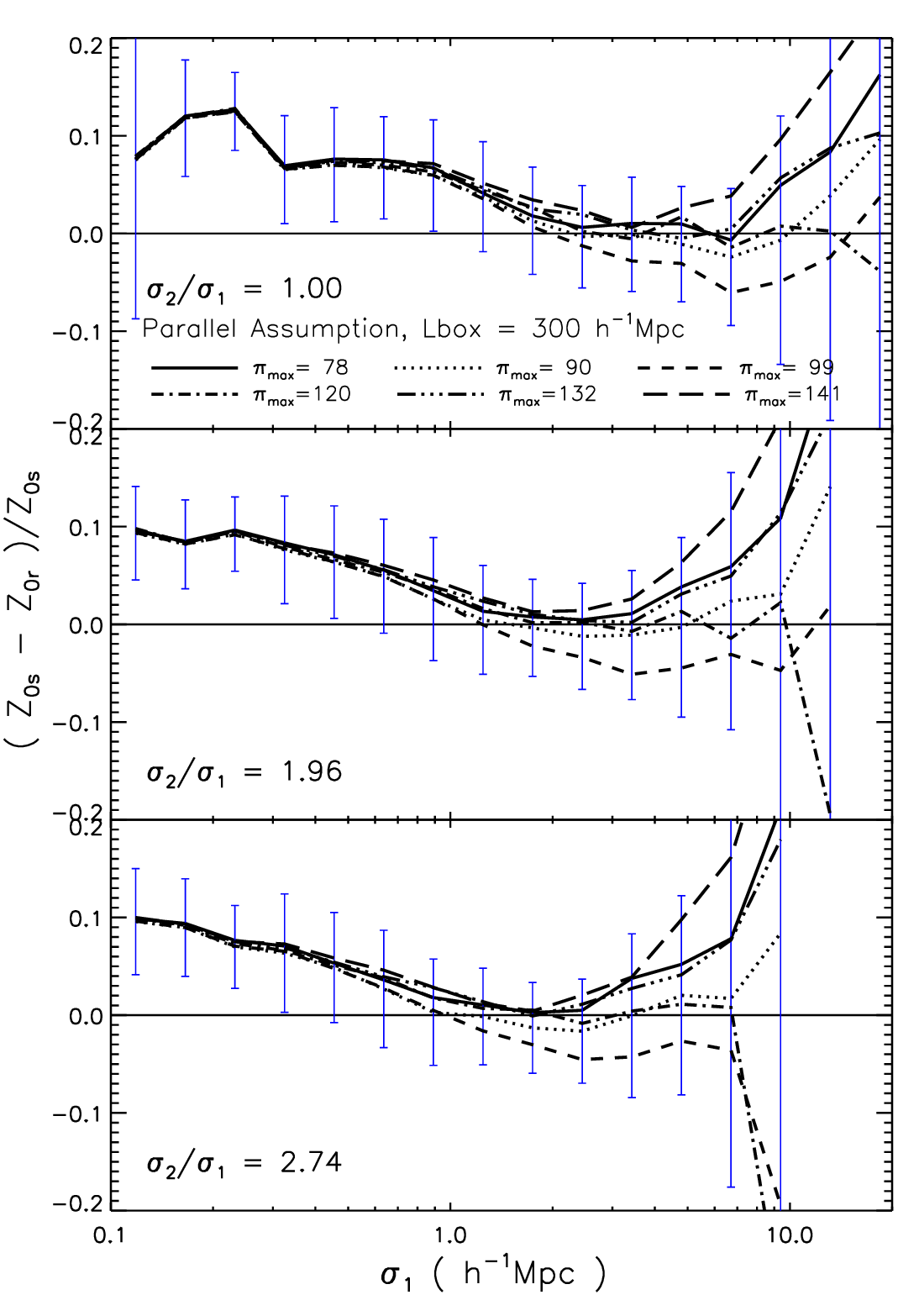}
\includegraphics{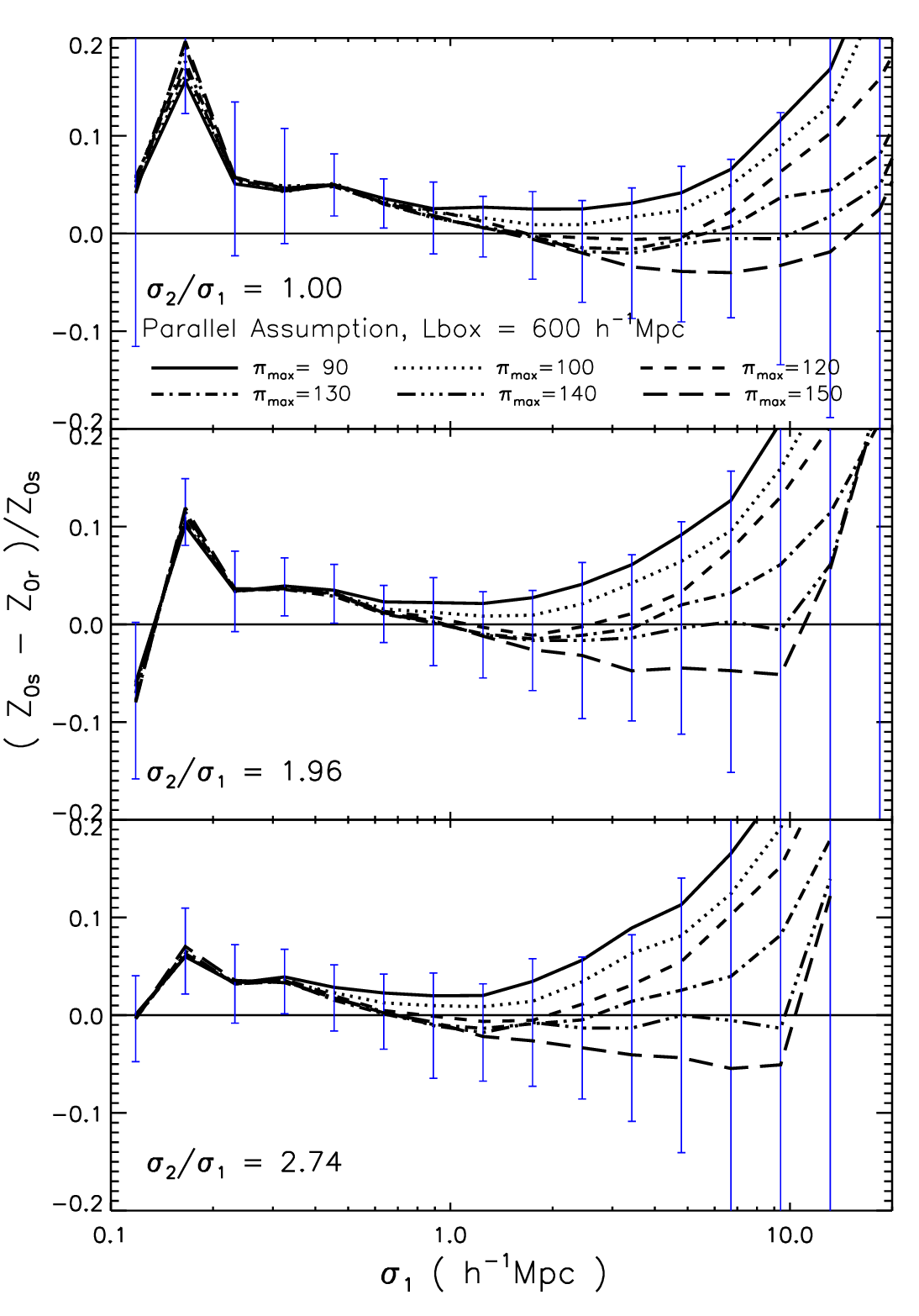}
}
\caption{Deviation of $Z_0$ in redshift space to $Z_0$ in real space
for different configurations of $\sigma_2/\sigma_1$ and choices of $\pi_{max}$. 
Left panel shows the wide angle treatment to the redshift distortion to the box300 
data, the middle shows the box300 results under plane parallel
assumption, and the right panel is shows box600 with plane parallel redshift distortion.}
\label{fig:z0rserror}
\end{figure*}

Figure~\ref{fig:z0rs} demonstrates $Z_0$ of the two simulation data sets in
redshift space and real space for different $\sigma_2/\sigma_1$ and $\pi_{max}$; 
detailed comparison is drawn in Figure~\ref{fig:z0rserror}. 
On most scales of $\sigma< 10h^{-1}$Mpc, residual effects of redshift distortion
due to finite integration domain result in only a minor bias within
$10\%$, except for an upshot in $Z_0$ in redshift space on scales of
$\sigma < \sim0.2 h^{-1}$Mpc. Adjusting $\pi_{max}$ does not modulate $Z_0$
significantly on $\sigma < \sim 1h^{-1}$Mpc, but causes some apparent deviations 
on larger scales, especially where $Z_0$ approaches its zero-crossing point. 
Nevertheless, it is reassuring from Figures~\ref{fig:z0rs} and \ref{fig:z0rserror}
that $Z_0$ estimated in redshift space agrees well with that of 
real space with at most $10\%$ uncertainty for $0.2< \sigma < \sim 10h^{-1}$Mpc
and $\pi_{max}\sim 120h^{-1}$Mpc. Thus $Z_0$ can be accepted as a redshift distortion 
free third order statistics to a good precision.

\section{halo model prediction of $Z_0$}

\subsection{Formalism}
The halo model invoked to model the third-order statistics $Z_0$ of dark matter
basically follows \citet{MaFry2000a, MaFry2000b}, \citet{ScoccimarroEtal2001}, 
\citet{FosalbaPanSzapudi2005} and \citet{SmithEtal2008}. 
Here we just give a brief description of main ingredients of the model, 
for more details we refer to the review of \citet{CooraySheth2002}.

\begin{enumerate}
\item Halo profile $\rho(r)$. It has been pointed out
that the density profile of a virialized dark matter halo in general is 
ellipsoidal and shows various morphology rather than a simple universal
spherical approximation \citep{JingSuto2000, JingSuto2002}. The 
non-spherical shape of halo can evidently affect the halo model prediction of the
clustering of dark matter on small scales where the one-halo term dominates 
\citep{SmithWattsSheth2006}. Noting that $Z_0$ is a degenerated 3PCF in 
analogous to $\zeta_0$, and should be similarly insensitive to halo shapes, 
the popular NFW profile \citep{NFW1997} is still adequate for our model.
For a halo of mass $M$ it reads,
\begin{equation}
u(r) = \frac{\rho(r)}{M}=\frac{fc^3}{4\pi R^3_{v}} \frac{1}{cr/R_v(1+cr/R_v)^2}\ ,
\label{eq:profile}
\end{equation}
where
$f = \left[ \ln{(1+c)}-c/(1+c) \right]^{-1}$ and 
$c(M) = c_0 (M/M_*)^{-\beta}$ 
known as the concentration parameter; parameters $c_0=9$, $\beta=0.13$ are calibrated 
by numerical simulation \citep{BullockEtal2001}. Halo mass is defined as 
$M=(4\pi R_v^3/3) \Delta \rho_{crit}$ with $\rho_{crit}$ being the cosmological critical
density. $\Delta$ is the density contrast for virialization and can be estimated
from spherical collapse model. A good fit for a flat universe with 
cosmological constant is given by \citet{BryanNorman1998}
\begin{equation}
\Delta=\frac{18\pi^2+83x-39x^2}{\Omega(a)}\ ,\ x=\Omega(a)-1\ .
\end{equation}

As it is more convenient to work in Fourier space
for 3PCF, the Fourier transformed halo profile of 
\citet{ScoccimarroEtal2001} is the used for 
computations
\begin{equation}
\begin{aligned}
u(k,M)= & f \bigg\{ \sin \eta \left[ \rm{Si} (\eta[1+\eta])-\rm{Si}(\eta)\right] \\
&+\cos \eta \left[ \rm{Ci} (\eta[1+\eta])-\rm{Ci}(\eta)\right] -
\frac{\sin{\eta c}}{\eta(1+c)}\bigg\}
\end{aligned}
\end{equation}
with $\eta=kR_v/c$.
Note that the halo profile is presumably truncated at 
virial radius $R_v$ in the standard version of halo model for large
scale structure; in extensions it becomes an adjustable 
parameter in an attempt to find the best match to simulations. 

\item Mass function $n(M)$. There are many versions of halo mass functions, 
but it turns out that using mass function with
higher precision actually brings about only a relatively minor change 
to $Z_0$ on the small 
scales as we tested. The classical Sheth-Tormen function \citep{ShethTormen1999} is 
sufficient,
\begin{equation}
n(M)M{\rm d}M 
=\bar{\rho} \frac{dy}{y} A \gamma \sqrt{\frac{g(\nu)}{2\pi}}
\left(1+g(\nu)^{-p} \right) \exp(-\frac{g(\nu)}{2})
\end{equation}
in which $\gamma={\rm d}\ln\sigma_M^2/{\rm d}\ln R$, $g(\nu)=\alpha \nu^2$, 
$\nu=\delta_c/\sigma_M$, $y=(M/M_*)^{1/3}=R/R_*$ with
$R_*=R_v\Delta$ and $\sigma_M$ being the extrapolated linear variance of the dark
matter fluctuations smoothed over the Lagrangian scale $R$. $A=0.322$
$\alpha=0.707$, $p=0.3$ are parameters fitted to simulations 
by \citet{JenkinsEtal2001}.

Note that the definition of halo mass in the Sheth-Tormen
function is $M=4\pi\bar{\rho} \Delta R_v^3/3$ with
$\bar{\rho}=2.78\times 10^{11} \Omega_m h^2 M_{\sun} {\rm Mpc}^{-3}$ 
being the dark matter density of the present universe, while the halo mass in
in the NFW profile is defined with the critical mass $\rho_{crit}=\bar{\rho}/\Omega_m$,
conversion between halo parameters of the two sets is given in \citet{SmithWatts2005}. 
\citet{ScoccimarroEtal2001} already noticed the inconsistency but argued
that effects of the difference could be largely cancelled in practical calculations,
and \citet{FosalbaPanSzapudi2005} also find that changing the 
concentration parameter by as much as $50\%$ would not affect the final 
results significantly. This is also the case in our calculation.

\item Halo bias. The distribution of massive halos is biased to the dark matter.
Most halo bias functions extracted from N-body simulations
\citep[e.g.][]{MoJingWhite1997, Jing1999, ShethTormen1999, TinkerEtal2010} 
are refinements to the analytical model of \citet{MoWhite1996}.
The bias plan used in our recipe is the fitting formula given 
by \citet{ShethTormen1999},
\begin{equation}\label{eq:bias}
b(\nu) = 1 + \frac{g(\nu)-1}{\delta_c} + \frac{2p}{\delta_c(1+g(\nu)^p)}\ ,
\end{equation}
in which $\delta_c$ is the linear overdensity threshold for spherical collapse. 
Its cosmological dependence is so weak that a constant value of $1.686$ is usually 
taken. The most recent update of \citet{TinkerEtal2010} is 
also applied in our code for a consistency check, and the results indicate that
the improvement to $Z_0$ is minor in the intermediate nonlinear regime only; it  
does not bring significant improvement to the overall accuracy when
considering the magnitude of numerical errors of estimation of the previous section.
\end{enumerate}

To prevent multi-dimensional integrations involved in direct calculation
of $\zeta$ in configuration space \citep{TakadaJain2003}, we work in 
Fourier space to yield the bispectrum $B$ predicted by halo model first. Then $B_0$ 
is easily obtained to render $Z_0$ through the transformation of Eq.~\ref{eq:b0z0}. 
In the halo model, bispectrum consists of three separate terms, namely the 
one-halo, two-halo and three-halo terms,
\begin{equation}
B(k_1, k_2, \phi) =B_{1h} +B_{2h} +B_{3h}\ ,
\end{equation}
in which
\begin{equation}
\begin{aligned}
B_{1h} & =I_{03}(k_1, k_2, k_3) \\
B_{2h} & =I_{11}(k_1)I_{12}(k_2, k_3)P_L(k_1)+ cyc.\\
B_{3h} & = \left[\prod_{i=1}^3  I_{11}(k_i)\right]B_{PT}
\end{aligned}
\end{equation}
and
\begin{equation}
I_{ij} =\int \, \frac{dr}{r}n(r)b_i(r)[u (k_1,r)\ldots u(k_j,r)]
\left( \frac{4\pi r^3}{3}\right)^{j-1}
\end{equation}
with $b_0=1$, $b_1=b(\nu)$ and $b_i=0$ for $i>1$ to neglect quadratic and high order
biasing terms. $P_L$ is the linear power spectrum and it is generated
by CMBFAST \citep{CMBFAST1996} with the cosmological 
parameters from the simulations we use. 
$B_{PT}$ is the bispectrum predicted by
the Eulerian perturbation theory at tree level \citep[e.g.][]{BernardeauEtal2002}.

\subsection{Comparison with simulations}

\begin{figure*}
\resizebox{\hsize}{!}{\includegraphics{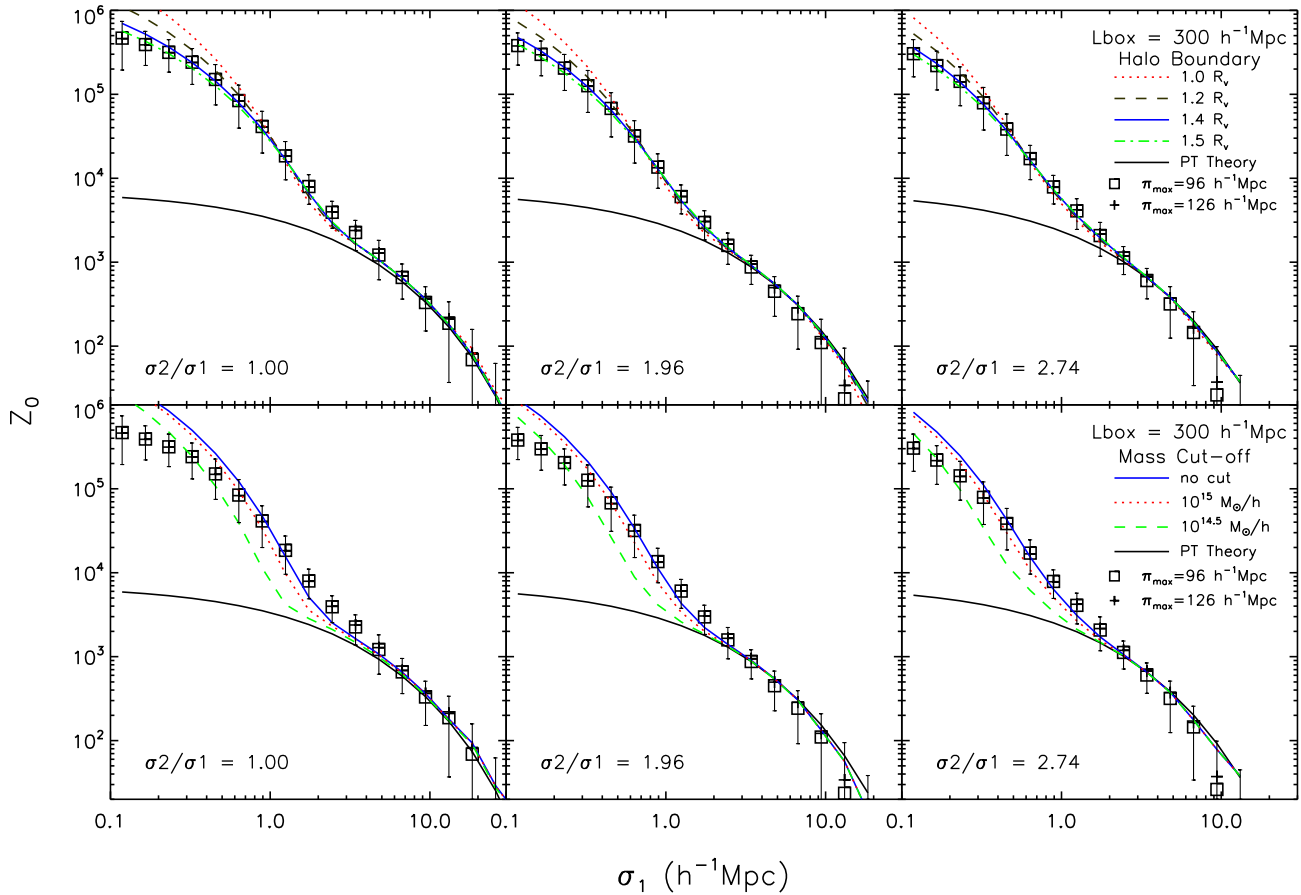}}
\caption{$Z_0$ of the box300 simulation (z=0) with predictions from halo model and
Eulerian perturbation theory (PT). Symbols are measurements of simulations in real space
with different $\pi_{max}$. The upper row of plots shows the 
effects of adjusting halo boundary radius
in unit of $R_v$ but without a cut-off to the halo mass function, while the bottom row
demonstrates the consequence of cutting the high mass tails of the mass function with
the halo boundary fixed to $R_v$.}
\label{fig:z0box300}
\end{figure*}

\begin{figure*}
\resizebox{\hsize}{!}{\includegraphics{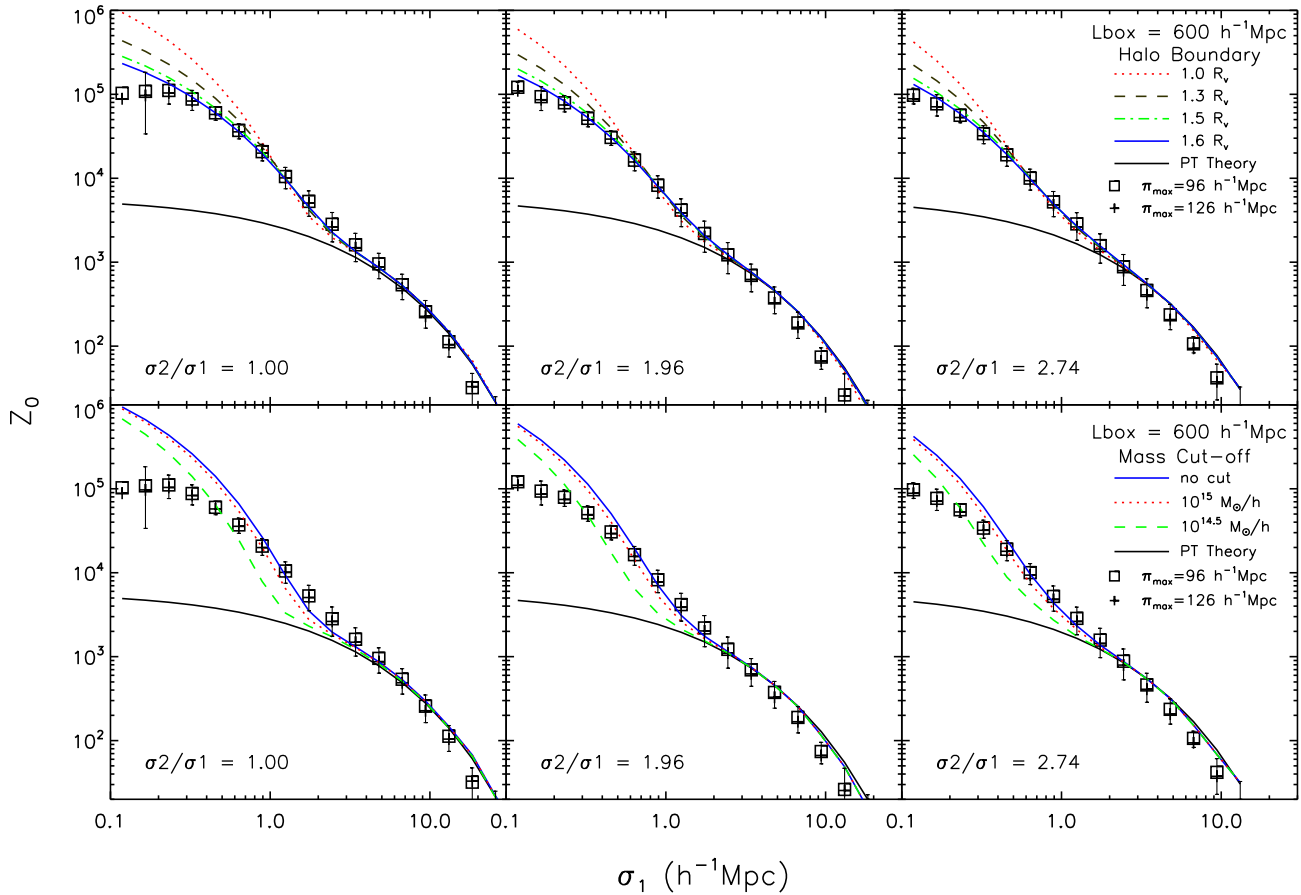}}
\caption{$Z_0$ of the box600 simulation data and models. See last figure for details.}
\label{fig:z0box600}
\end{figure*}

\begin{figure}
\resizebox{\hsize}{!}{\includegraphics{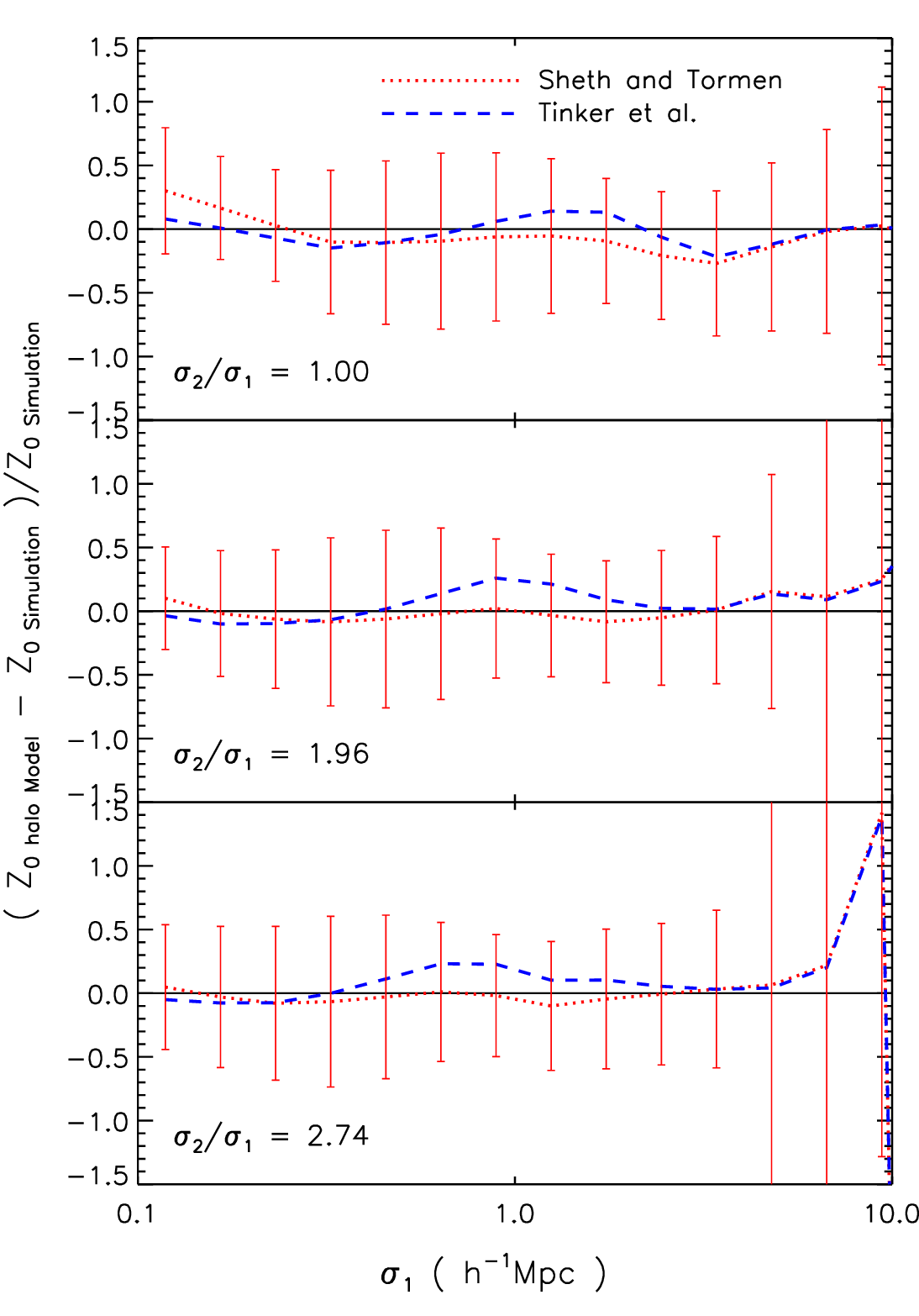}}
\caption{Relative differences of the halo model predicted $Z_0$ to that of 
estimated from the box300 simulation. Red dotted lines correspond to the 
prediction of the halo model
with halo mass function
and bias function given by \citet{ShethTormen1999}, blue dashed lines are
generated by the model with mass function and bias function provided
by \citet{TinkerEtal2010}. $Z_{0\, simulation}$ is the estimation from 
the box300 data with plane-parallel assumption to the redshift distortions.}
\label{fig:haloerror}
\end{figure}

$Z_0$ predicted by halo model and Eulerian perturbation theory is 
demonstrated in Figures~\ref{fig:z0box300} and \ref{fig:z0box600} overlaid with
measurements of the box300 and the box600 simulation data, respectively,
estimated in real space.
Results of our halo model and simulations agree remarkably well
at both redshifts $z=0, 0.1$, especially on $\sigma_1$ scales
between $\sim 0.2 - \sim 5h^{-1}$Mpc. 

On very small scales $\sigma_1 < \sim 0.2h^{-1}$Mpc, the halo model 
predicted $Z_0$ is larger and steeper
than the simulation. This is more apparent for box600. Numerical
tests reveal that this is partly due to dilution to the original 
data set: a higher density of points leads a higher clustering power 
in this regime. On large scales, where halo model follows perturbation theory, 
both theories begin to over-predict the clustering strength of simulations
for larger $\sigma_2/\sigma_1$, which should not be attributed to the imperfection
of halo models and should be the inaccuracy of $B_{PT}$ on these 
scales \citep{PanColesSzapudi2007, GuoJing2009}.

To improve halo model performance at the three-point level,
halo boundary and mass function adjustments are usually 
adopted \citep{TakadaJain2003, WangEtal2004, FosalbaPanSzapudi2005}. This
alleviates the disagreement to some extent. Here we also
enlarge halo boundary beyond $R_v$ and truncate
the high mass tail of halo mass function 
(Figures~\ref{fig:z0box300} and \ref{fig:z0box600}).
Experiment indicates that this extension of halo radius without a hard cut-off of
the mass function can easily generate the correct shape and amplitude 
of $Z_0$ of simulations.
Simple fitting shows that best halo boundary is $\sim1.5R_v$ for box300 and 
$\sim1.6R_v$ for box600. In contrast,
if we keep the halo boundary unchanged but truncate the halo mass function,
the one- and two-halo terms are so strongly modified, and the shape and the 
amplitude of $Z_0$ deviate from simulations significantly. 
Simple fitting to simulations by setting both halo boundary and mass cut-off free
reveals that mass cut-off could not be smaller than $10^{15}M_{\sun}$. 
Otherwise,
there is no way to reconcile the under-predicted $Z_0$ with simulations at transition scales
of $\sigma_1\sim 3h^{-1}$Mpc, above which $B_{PT}$ breaks. 
In conclusion, enlarging halo boundary alone is sufficient for 
for accurately predicting $Z_0$.

During our calculation, we also examine the influence of the halo bias function and the
mass function by using the high precision 
formula of \citet{TinkerEtal2010}. Such replacement
does not cause a fundamental change to the theoretical 
prediction (Figure~\ref{fig:haloerror}). 
The new mass functions do not benefit the halo model much. On most $\sigma$ scales, 
less than $10^{-1}$Mpc, the halo model with Sheth-Tormen
functions is consistent with simulations within our error budget of $\sim10-20\%$. 
The replacement of functions provided by \citet{TinkerEtal2010} increases
deviation level to around $20-30\%$, especially on scales of $\sim 1h^{-1}$Mpc;
visible advantage only just appears on scales of $\sigma_1> \sim3h^{-1}$Mpc 
with accuracy gain of a few percents.

In addition to the halo model, we also checked the phenomenological models 
of \citet{ScoccimarroCouchman2001} and \citet{PanColesSzapudi2007}. 
The accuracy of the formula by \citet{ScoccimarroCouchman2001} is
very good on scales $\sigma_1 > \sim 1h^{-1}$Mpc but then deviates from
the simulations
by more than 40\% on smaller scales. The performance of \citet{PanColesSzapudi2007} is
poor in terms of $Z_0$ as the bispectrum 
model is not designed to conserve clustering power
and the resulting integration over it yields incorrect amplitude. Nevertheless,
if a renormalization is enforced for the model 
to be consistent with the perturbation theory 
on large $\sigma$, the model
works well for $Z_0$ at $\sigma_1> \sim 2h^{-1}$Mpc.

\section{Summary and Discussion}
In this paper we propose a third-order correlation function for characterising
galaxy clustering properties. The statistics $Z_0$ we advocate
is the zeroth-order component of the projected 3PCF.
Although $Z_0$ is a 3PCF, its estimation takes roughly the same 
amount of computing operation as the projected 2PCF. The algorithm
can be easily implemented after moderate modification of a code for 
the projected 2PCF.

Various numerical experiments confirm that $Z_0$ can be deemed to be
redshift distortion free within approximately $10\%$
for the regime where the scale 
perpendicular to LOS is $0.2<\sigma<10h^{-1}$Mpc. In addition, the maximal
integration scale $\pi_{max}$ parallel to LOS during estimation ought to be 
greater than $\sim 120h^{-1}$Mpc. A serious concern is that shot noise could ruin
the estimation in the strongly nonlinear regime if the number density of points 
in a sample is too low. This requirement for a robust $Z_0$ measurement is
tighter than for the projected 2PCF, but still weaker than the normal projected 3PCF,
since $Z_0$ is an integral of the former. The criterion we suggest is $DDD>\sim 100$.

As we expected, the halo model provides satisfactory prediction to dark matter
$Z_0$ of simulations within $\sim10\%$, if the classical Sheth-Tormen mass
functions are used. Our computation indicates that extending the halo
boundary is enough to yield good fit to simulations, while a hard cut-off to mass function
is not as effective as previous works claimed. Substituting new 
functions of the halo mass distribution and halo biasing in high precision 
does not lead to significantly better agreement with simulations. 
Since the angular dependence 
in the projected 3PCF and the normal 3PCF is smeared out in $Z_0$, 
we conjecture that using an anisotropic halo profile probably will 
not significantly improve accuracy.
A significant bias of halo model predicted
$Z_0$ compared to simulations emerges in the weakly nonlinear regime, where
halo models boil down to second-order perturbation
theory; the latter is 
already known to be poor in predicting dark matter 3PCF. A
more precise bispectrum from higher order perturbation 
theories may offer a way to increase precision
\citep[e.g.][]{Valageas2008, Sefusatti2009, BartoloEtal2010}.

The principal reason for proposing $Z_0$ is to provide an efficient 
redshift distortion free 3PCF, complementary to the standard projected 2PCF,
for galaxy clustering analyses. It is well known that
the projected 2PCF itself is a Gaussian statistic only and thus has its
limitations. Third order correlation functions, mainly carrying
information about non-Gaussianity, are more sensitive to
details of the galaxy distribution. Non-Gaussianity of galaxy distribution is 
generated by the nonlinear
action of gravitational force and gas physics if the primordial density 
fluctuation of the universe after inflation is Gaussian. The degeneracy shown
in projected 2PCF \citep[e.g.][]{ZuEtal2008} may be broken if third
order correlation functions are employed.
The redshift distortion free feature of $Z_0$ on scales less than $10h^{-1}$Mpc
defines its potential in investigating the relation of galaxies with their host halos, 
and the formation histories of galaxies and halos. 
Furthermore, the success of halo model prediction on dark matter $Z_0$ 
encourages us to apply $Z_0$ for analysing galaxies. In principle, with measurements 
from galaxy samples, $Z_0$ enables us to 
generalize and diagnose schemes of HOD, conditional luminosity 
function \citep[CLF, ][]{YangEtal2003} and semi-analytical models 
\citep[e.g.][]{Baugh2006} to 
third order statistics at cost of one additional free
parameter, the halo boundary. Our present work is restricted to dark matter only,
the behavior of $Z_0$ for biased objects remains unclear. Testing with mock galaxy
samples before applying to real data will be necessary.

\section*{Acknowledgment}
This work is supported by the Ministry of Science \& Technology of China
through 973 grant of No. 2007CB815402 and the NSFC through
grants of Nos. 10633040, 10873035.
JP acknowledges the One-Hundred-Talent fellowship of CAS. IS
acknowledges support from NASA grants NNG06GE71G and NNX10AD53G and
from the Pol{\'a}nyi Program of the Hungarian National Office for
Research and Technology (NKTH).
We thank Weipeng Lin for his kindness of providing us N-body
simulation data.

\end{document}